# First-principles Predictions of Electronic Properties of GaAs$_{1-x-y}$P$_y$Bi$_x$ and GaAs$_{1-x-y}$P$_y$Bi$_x$–based Heterojunctions

Guangfu Luo,[1] Kamran Forghani,[2] Thomas F. Kuech,[2] Dane Morgan[1],*

[1]*Department of Materials Science and Engineering, University of Wisconsin-Madison, Madison, Wisconsin 53706, USA*

[2]*Department of Chemical and Biological Engineering, University of Wisconsin-Madison, Madison, Wisconsin 53706, USA*

*Author to whom correspondence should be addressed. E-mail: ddmorgan@wisc.edu

Significant efficiency droop is a major concern for light-emitting diodes and laser diodes operating at high current density. Recent study has suggested that heavily Bi-alloyed GaAs can decrease the non-radiative Auger recombination and therefore alleviate the efficiency droop. Using density functional theory, we studied a newly fabricated quaternary alloy, GaAs$_{1-x-y}$P$_y$Bi$_x$, which can host significant amounts of Bi, through calculations of its band gap, spin-orbit splitting, and band offsets with GaAs. We found that the band gap changes of GaAs$_{1-x-y}$P$_y$Bi$_x$ relative to GaAs are determined mainly by the local structural changes around P and Bi atoms rather than their electronic structure differences. To obtain alloy with lower Auger recombination than GaAs bulk, we identified the necessary constraints on the compositions of P and Bi. Finally, we demonstrated that GaAs/GaAs$_{1-x-y}$P$_y$Bi$_x$ heterojunctions with potentially low Auger recombination can exhibit small lattice mismatch and large enough band offsets for strong carrier confinement. This work shows that the electronic properties of GaAs$_{1-x-y}$P$_y$Bi$_x$ are potentially suitable for high-energy infrared light-emitting diodes and laser diodes with improved efficiency.

## Introduction

High-power infrared light-emitting diodes (LEDs) and laser diodes (LDs) have important applications in a number of areas, such as telecommunication, material processing, medical applications, and military defense. However, the efficiency of LEDs and LDs operating at high current densities (over ~30 A/cm$^2$ for LEDs[1,2] or over ~1 kA/cm$^2$ for LDs[3,4]) dramatically decreases, a phenomenon known as efficiency droop.[5,6] To achieve high-power output, the usual remedy is employing parallel diode arrays operating around an optimal current density. This method, however, increases the volume, complexity, and cost of devices, and in turn prohibits the more widespread applications of high-power infrared LEDs and LDs.

The efficiency droop can be induced by three major sources: defects, electron leakage, and Auger recombination.[5-7] Recent experiments have confirmed that Auger recombination is the major cause of the efficiency droop in InGaN/GaN LEDs,[8] a finding possibly true for other similar systems. Because direct Auger recombination is roughly proportional to $n^3 e^{-CE_g/k_B T}$,[7] where $n$ is carrier density, $E_g$ the band gap, and $C$ a material-dependent coefficient, this mechanism is especially significant for devices operating at high current density and with small band gap, as confirmed in a number of different materials.[5,9] For GaAs-based infrared diodes, it has been proposed[10,11] to both decrease the band gap and increase the spin-orbit (SO) splitting by Bi alloying to reduce one type of Auger recombination process, CHSH, where a Conduction electron recombines with a Heavy hole and the released energy subsequently excites a Spin-split-off valence electron into a Heavy hole band. A recent estimation[12] indicated that the rate of the CHSH process is 35% to 23% of the total Auger recombination rate, when the band gap of GaAs-based alloy changes from 1.4 to 0.7 eV, respectively.

However, it has been difficult to grow the ternary alloy GaAs$_{1-x}$Bi$_x$ with good structural[13,14] and optical[15] properties because of the various defects induced by the large strain around Bi atoms.[16,17] Recently, we fabricated a quaternary alloy GaAs$_{1-x-y}$P$_y$Bi$_x$ with P fraction up to 30.6% and Bi fraction up to 8.5% using metal-organic vapor phase epitaxy.[18] Through co-incorporation of P and Bi into GaAs, it is feasible to enhance the solubility of Bi because of the relatively small atom size of P and meanwhile to tune the lattice and electronic properties. In this letter, we present our theoretical investigations of two essential issues associated with applying this alloy to the active layer of infrared LEDs and LDs. The first issue is the band gap and SO splitting of GaAs$_{1-x-y}$P$_y$Bi$_x$, which determines the Auger recombination. The second issue is the band offsets of the GaAs/GaAs$_{1-x-y}$P$_y$Bi$_x$ heterojunction, a property determining the carrier confinement in the double heterojunction GaAs/GaAs$_{1-x-y}$P$_y$Bi$_x$/GaAs, which confinement is critical to obtain high internal quantum efficiencies.

**Theoretical Details**

We carry out all calculations using density functional theory (DFT) as implemented in the Vienna *Ab initio* Simulation Package.[19,20] The local density approximation (LDA) functional is employed and SO coupling is included to describe the important special relativity effects in Bi atoms. The projector augmented wave method is used with the following potentials: Ga_GW (4s$^2$4p$^1$), As_GW (4s$^2$4p$^3$), P_GW (3s$^2$3p$^3$), and Bi_d_GW (5d$^{10}$6s$^2$6p$^3$). The plane-wave energy cutoff is set to 400 eV. The supercells in band structure and band offset calculations consist of 3×3×3 and 2×2×6 of the GaAs conventional cells, including 216 and 192 atoms or a volume of ~16.9×16.9×16.9 and ~11.2×11.2×33.9 Å$^3$, respectively. In band offset calculations, a heterojunction is constructed along [001] and the two different materials have equal number of atoms. We predict a binding energy of 11 meV for P and Bi on neighboring As sites relative to isolated P and Bi dopants, suggesting that the P and Bi dopants form a nearly ideal solid solution at relevant synthesis temperatures, where $k_BT$ is much larger than 11 meV. Thus we randomly distribute P and Bi atoms on the As sites using the special quasirandom structure (SQS) approach. The SQS with pair and triplet clusters within a radius of 5.8 Å is optimized with a Monte Carlo algorithm, as implemented in the ATAT code.[21] The Monkhorst-Pack *k*-point grids for Brillouin zone sampling are 3 × 3 × 3 and 6 × 6 × 1 for the 216 and 192-atom structures, respectively. The lattice length is relaxed either in all directions for unstrained structures or only in the [001] direction for structures with (001) plane strained to GaAs(001). The remaining force on each atom is less than 0.01 eV/Å after relaxation.

The band gap underestimation by LDA functional is fixed by operating on the LDA gap with a linear function: 0.88 eV + 1.13 $E^g$, where $E^g$ is the LDA band gap. This correction gives results in good agreement with previous experimental band gaps (see supplementary material). Band offsets of the GaAs/GaAs$_{1-x-y}$P$_y$Bi$_x$ heterojunction are calculated based on positions of valance band maximum (VBM) and conduction band minimum (CBM) relative to macroscopic electrostatic potential (MEP)[22] with a procedure detailed in the supplementary material.

**Result and Discussion**

**Band gap of GaAs$_{1-x-y}$P$_y$Bi$_x$ bulk.** We investigate 25 compositions of GaAs$_{1-x-y}$P$_y$Bi$_x$, with $0 \leq x \leq 0.10$ and $0 \leq y \leq 0.40$ (see supplementary material), and consider both fully relaxed structures and ones strained to the GaAs(001). The DFT-based band gaps are fitted with Taylor series for easy calculations of other unexamined compositions, a method used also in the band gap study of GaAsNBi.[23] The fitting result of fully relaxed GaAs$_{1-x-y}$P$_y$Bi$_x$ is shown in Eqn. 1.

$$E^g_{GaAs_{1-x-y}P_yBi_x} = E^g_{GaAs_{1-x}Bi_x} + E^g_{GaAs_{1-y}P_y} - E^g_{GaAs} + c_5(E^g_{GaAs_{1-x}Bi_x} - E^g_{GaAs})(E^g_{GaAs_{1-y}P_y} - E^g_{GaAs}) \quad (1a)$$

$$E^g_{GaAs_{1-x}Bi_x} = E^g_{GaAs} + c_1 x + c_2 x^2 \quad (1b)$$

$$E^g_{GaAs_{1-y}P_y} = E^g_{GaAs} + c_3 y + c_4 y^2 \quad (1c)$$

$$E^g_{GaAs} = 1.42 \quad (1d)$$

$$c_1 = -7.32, \; c_2 = 15.10, \; c_3 = 1.16, \; c_4 = 0.33, \; c_5 = 0.11 \quad (1e)$$

The fitting coefficients for GaAs$_{1-x-y}$P$_y$Bi$_x$ strained to GaAs(001) are shown in Eqn. 2. All the band gaps in Eqn. 1 and 2 have a unit of eV. The standard deviations of the fittings are about 0.03 eV (see complete DFT-based band gaps and fitting error in the supplementary material).

$$c_1 = -6.51, \; c_2 = 11.67, \; c_3 = 0.80, \; c_4 = -0.12, \; c_5 = -0.62 \quad (2)$$

The contour plots for the band gap of fully relaxed and strained structures are shown in Fig. 1a and 1b, respectively. With increasing alloying content, P increases the band gap but Bi greatly reduces it. The relatively straight contour lines indicate that P and Bi atoms have weak cooperative effects to the band gap. For the strained case, the quaternary alloy possesses band gap less than that of GaAs (1.42 eV) under the condition $y < 8.10 x - 37.55 x^2 + 159.22 x^3$. The lattice-match condition is $y = 4.35 x$ according to Vegard's law and the calculated lattice lengths of GaP (5.47 Å), GaAs (5.62 Å), and GaBi (6.27 Å) bulks. As we will show in the next paragraph, such Vegard's lattice-match condition is different from the one where the fully relaxed and strained structures have the same band gaps.

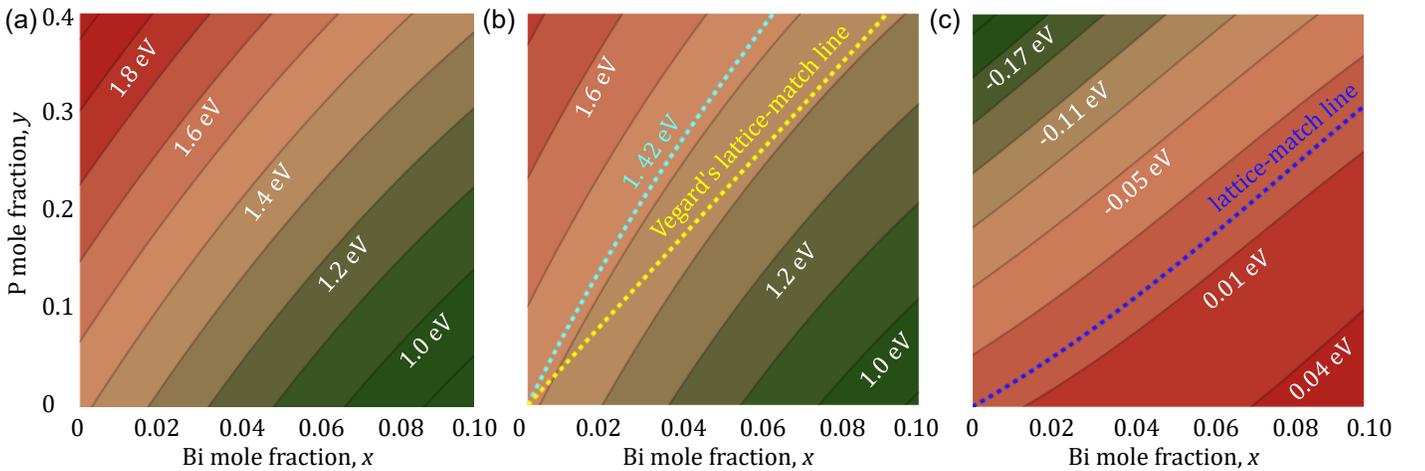

FIG. 1. Band gap of (a) fully relaxed and (b) strained GaAs$_{1-x-y}$P$_y$Bi$_x$. (c) Band gap differences of the strained alloy relative to its fully relaxed counterpart. The blue dashed line corresponds to the 0-eV contour line and is expected to be the real lattice-match line.

Figure 1c shows that the band gap difference between the strained and fully relaxed structures increases approximately linearly when the in-plane strain changes from tension to compression. Under the condition $y = 2.26\,x + 10.34\,x^2 + 9.91\,x^3 - 321.51\,x^4$ (blue dashed line), the stained structures have the same band gaps as the fully-relaxed counterpart, which means that the former are lattice-matched with GaAs(001). Such lattice-match condition is close to the one ($y = 2.66\,x + 1.87\,x^2 + 0.91\,x^3$) where the lattice lengths of fully relaxed structures equate with those of GaAs, and is expected to be more accurate than the one based on Vegard's law, which is disobeyed in a number of materials.[24] We therefore consider the lattice-match line in Fig. 1c as a reasonable lattice-match condition. The band gap change by the stain effect is -0.21–0.05 eV in the examined composition range. Detailed analyses based on mechanical properties and DFT-based band gaps (see supplementary material) show that the strain effect on the band gap can be described by Eqn. 3

$$\Delta E_g / E_g \approx -6.96\, \Delta a/a \qquad (3)$$

where $E_g$, and $a$ are the band gap and lattice parameter of the fully relaxed structure, and $\Delta E_g$ and $\Delta a$ are the band gap and in-plane lattice parameter changes of the strained alloy relative to its fully relaxed counterpart, respectively; the standard error of the coefficient is 0.24. With Eqn. 3, the band gap change induced by the strain of other substrates can be easily evaluated.

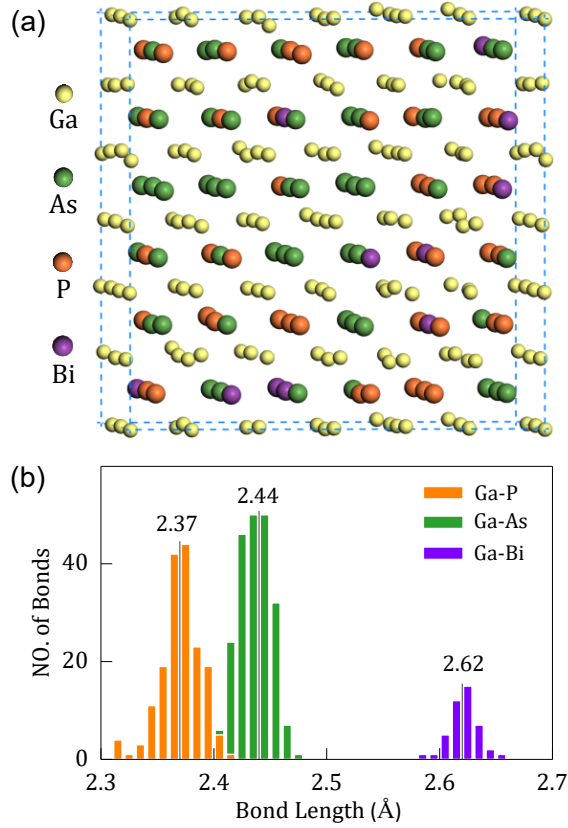

FIG. 2. (a) Optimized structure and (b) histogram of bond length in fully relaxed GaAs$_{0.5}$P$_{0.398}$Bi$_{0.102}$. Strained GaAs$_{0.5}$P$_{0.398}$Bi$_{0.102}$ shows very similar results. The values in panel (b) indicate the average lengths of the three types of bond.

To understand how the atomic radius and chemistry of P and Bi atoms influence the band gap change, we first examine the geometrical distortions of a heavily alloyed structure, GaAs$_{0.5}$P$_{0.398}$Bi$_{0.102}$. Figure 2a shows that local distortions are larger in the Ga layer than in the group-V layer, with an average displacement of 0.19 Å in the former relative to 0.09 Å in the latter. This phenomenon is related to the tetrahedral coordination in GaAs bulk: all defects on the As sites have locally centrosymmetric strain and thus stay near the As sites, but they push/pull four nearby Ga atoms. Statistics of the chemical bonds in Fig. 2b shows that the average lengths of the Ga-P, Ga-As, and Ga-Bi bonds is ~2.37, ~2.44, and ~2.62 Å, respectively. Note that the Ga-As bond length is 2.43 Å in GaAs bulk. Therefore, the major structural changes can be summarized as follows: P atoms slightly pull Ga atoms close but Bi markedly pushes Ga atoms away, a phenomenon consistent with the order of covalent radii of P (1.07 Å), As (1.19 Å) and Bi (1.48 Å).[25]

Then we separately assess two influences associated with the incorporation, the change in chemistry and the relaxation of the atomic positions, to understand the origin of band gap change induced by alloying. We choose $GaAs_{0.805}P_{0.093}Bi_{0.102}$ as an example, because it possesses band gap (0.93 eV) significantly different from that of GaAs bulk. First, substitution of all the P and Bi atoms in the fully relaxed $GaAs_{0.805}P_{0.093}Bi_{0.102}$ with As atoms generates a band gap of 1.08 eV, close to the 0.93 eV band gap of the optimized structure. Second, shift of the atom positions in fully relaxed $GaAs_{0.805}P_{0.093}Bi_{0.102}$ back to the reduced coordinates of GaAs bulk generates a band gap of 1.31 eV, very close to the 1.34 eV band gap of GaAs bulk. The two results indicate that the band gap changes in $GaAs_{1-x-y}P_yBi_x$ are mainly caused by the geometrical changes induced by P and Bi atoms, and the chemical and electronegativity differences among As, P, and Bi atoms are not critical. This result is different from the foundation of the widely-used valence band anti-crossing (VBAC) model in Bi-containing III-V alloys, where the model totally neglects structural distortions and ascribes the band gap changes to interactions between the valence bands of host semiconductor and the states of the alloying elements.[26] One prediction of the VBAC model is that the Bi-related states are localized near the valance-band edge.[26] This prediction, however, is in contradiction with our projected density of states results, which show that the Bi states are delocalized across the whole band structure, similar to the As states (see supplementary material). Therefore, a fitted VBAC model should be interpreted as an empirical way of describing primarily the effects of structural distortions in $GaAs_{1-x-y}P_yBi_x$.

**SO splitting of $GaAs_{1-x-y}P_yBi_x$ bulk.** Similar to the band gap of $GaAs_{1-x-y}P_yBi_x$, we also fit the DFT SO splitting of fully relaxed structures with a Taylor series as expressed in Eqn. 4.

$$E^{SO}_{GaAs_{1-x-y}P_yBi_x} = E^{SO}_{GaAs_{1-x}Bi_x} + E^{SO}_{GaAs_{1-y}P_y} - E^{SO}_{GaAs} + c_5(E^{SO}_{GaAs_{1-x}Bi_x} - E^{SO}_{GaAs})(E^{SO}_{GaAs_{1-y}P_y} - E^{SO}_{GaAs}) \quad (4a)$$

$$E^{SO}_{GaAs_{1-x}Bi_x} = E^{SO}_{GaAs} + c_1 x + c_2 x^2 \quad (4b)$$

$$E^{SO}_{GaAs_{1-y}P_y} = E^{SO}_{GaAs} + c_3 y + c_4 y^2 \quad (4c)$$

$$E^{SO}_{GaAs} = 0.35 \quad (4d)$$

$$c_1 = 4.11, c_2 = -6.91, c_3 = -0.14, c_4 = -0.17, c_5 = -2.32 \quad (4e)$$

The fitting coefficients for strained $GaAs_{1-x-y}P_yBi_x$ are shown in Eqn. 5. All the band gaps in Eqn. 4 and 5 have a unit of eV. The standard deviations of the above fittings are about 0.01 eV (see complete DFT SO splitting and fitting error in the supplementary material).

$$c_1 = 3.99, c_2 = 0.83, c_3 = -0.05, c_4 = 0.07, c_5 = 18.66 \quad (5)$$

The contour plots for the SO splitting of fully relaxed and strained structures are shown in Fig. 3a and 3b, respectively. As expected, P atoms decrease the SO splitting but Bi atoms increase it. Different from the strain effects on band gap, both strong tensile and compressive strains enhance the SO spitting (Fig. 3c), although the changes are less than ~0.07 eV in the examined composition range. To obtain SO splitting greater than that of GaAs (0.35 eV), the composition should satisfy $y < 74.38\ x$. For GaAs$_{1-x-y}$P$_y$Bi$_x$ with lower Auger recombination ratio than GaAs, it should possess a smaller band gap ($y < 8.10\ x - 37.55\ x^2 + 159.22\ x^3$) and a larger SO splitting ($y < 74.38\ x$) than GaAs, which therefore sets up a composition requirement of at least $y < 8.10\ x - 37.55\ x^2 + 159.22\ x^3$.

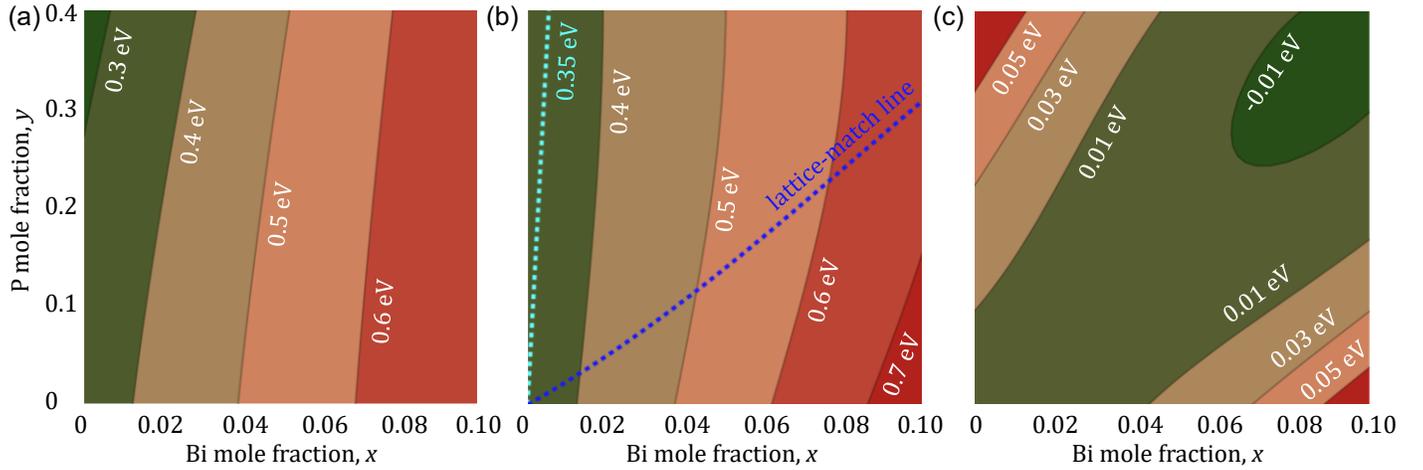

FIG. 3. SO splitting of (a) fully relaxed and (b) strained GaAs$_{1-x-y}$P$_y$Bi$_x$. The 0.35-eV contour line can be approximated as $y = 74.38\ x$. The lattice-match line is same as the one in Fig. 1c. (c) SO splitting differences of the strained alloy relative to its fully relaxed counterpart.

**Comparison between theoretical predictions and experiments.** To verify the abovementioned results, we compare our predictions of the band gap and SO splitting with available experiments. Our theoretical results are generally in good agreement with the experimental band gap[27] of GaAs$_{1-y}$P$_y$ and band gap[11,28-30] and SO splitting[11,30,31] of GaAs$_{1-x}$Bi$_x$ (see Fig. S4 in supplementary material). In the examined ranges, the largest differences between the experimental data or best fitting of the experimental data and Eqn. 1b, Eqn. 1c, Eqn. 4b are about 0.03, 0.02, and 0.08 eV, respectively. The agreement is also reasonably good with the limited experimental band gap[18] of GaAs$_{1-x-y}$P$_y$Bi$_x$. According to Eqn. 2, the band gaps of strained structure with $(x, y)$ = (0.031, 0.280), (0.070, 0.273), and (0.085, 0.230) are 1.47, 1.28, 1.18 eV, which have an average deviation of about 0.08 eV relative to the experimental values of 1.39, 1.19, and 1.11 eV, respectively.

**Band offset of GaAs/GaAs$_{1-x-y}$P$_y$Bi$_x$ heterojunction.** To examine if efficient carrier confinement can be realized in GaAs$_{1-x-y}$P$_y$Bi$_x$-based LEDs and LDs, we investigate the band offsets of GaAs/GaAs$_{1-x-y}$P$_y$Bi$_x$ heterojunctions. Figure 4 compares the band diagrams of three heterojunctions with 8.3% Bi and three P contents: GaAs/GaAs$_{0.917}$Bi$_{0.083}$, GaAs/GaAs$_{0.771}$P$_{0.146}$Bi$_{0.083}$, and GaAs/GaAs$_{0.625}$P$_{0.292}$Bi$_{0.083}$, with respective lattice mismatch of 0.99%, 0.65%, and 0.31% based on our calculated lattice parameters of fully-relaxed GaAs and GaAs$_{1-x-y}$P$_y$Bi$_x$. All the three quaternary alloys satisfy the composition requirement of having less Auger recombination than GaAs bulk ($y < 8.10\ x - 37.55\ x^2 + 159.22\ x^3$). Figure 4a and 4b show that GaAs/GaAs$_{0.917}$Bi$_{0.083}$ and GaAs/GaAs$_{0.771}$P$_{0.146}$Bi$_{0.083}$ have a type-I (straddling gap) band diagram, which is favorable for optical transitions, and the VBM (CBM) band offsets are 240 (220) and 170 (140) meV for the former and latter heterojunctions, respectively. Because the band offsets of the two heterojunctions are 5.4–9.2 times of the thermal energy at room temperature (~26 meV at 300 K), they are expected to exhibit effective carrier confinement and low leakage current at temperature $T$.[32] However, Fig. 4c shows that GaAs/GaAsP$_{0.292}$Bi$_{0.083}$ heterojunction owns the type-II (staggered gap) band diagram with a VBM (CBM) band offset of 190 (20) meV, and thus is less useful for utilization in optical transitions than the previous two. These results demonstrate that by carefully choosing the composition of GaAs$_{1-x-y}$P$_y$Bi$_x$ it is possible to obtain heterojunctions that simultaneously own large band offsets, small lattice mismatch, and band gap and SO splitting resulting in reduced Auger recombination.

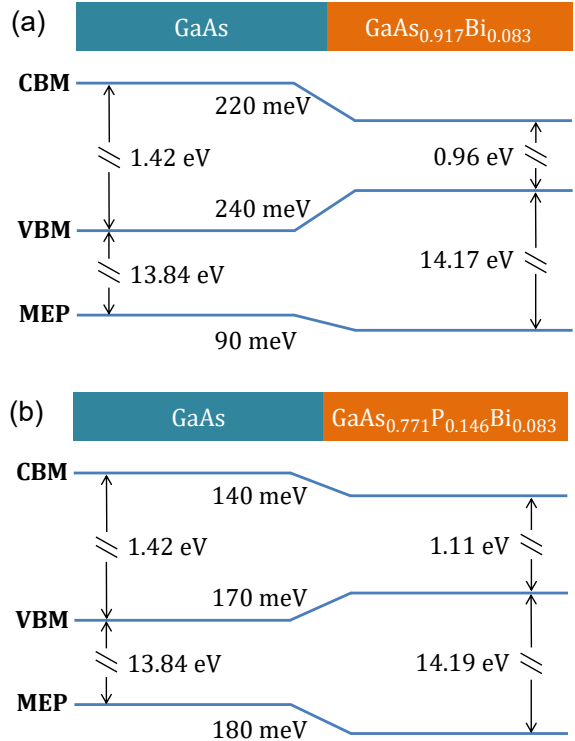

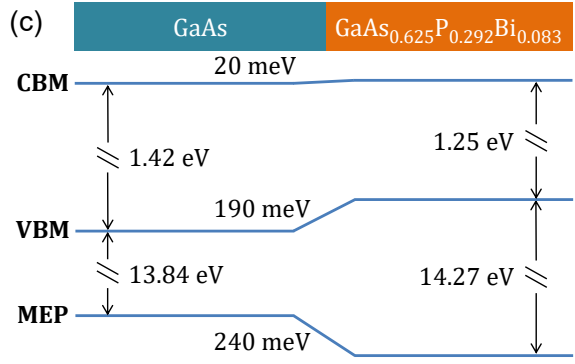

FIG. 4. Band diagrams of heterojunctions formed between GaAs and (a) GaAs$_{0.917}$Bi$_{0.083}$, (b) GaAs$_{0.771}$P$_{0.146}$Bi$_{0.083}$, and (c) GaAs$_{0.625}$P$_{0.292}$Bi$_{0.083}$. The band offsets, band gaps, and positions of VBM relative to MEP are indicated. The transitional region near the interface is found to be ~2.8 Å.

It is worthwhile to point out that the previous estimation[33] of band offsets based on band structures of the two bulk materials forming the heterojunction is likely unreliable for GaAs$_{1-x-y}$P$_y$Bi$_x$. This is because the interface of a heterojunction usually induces electric dipoles, which lead to the MEP offset.[34] For the three examined heterojunctions, the MEP offsets are 90–240 meV, which are comparable to or even greater than the band offsets. The estimated band offsets based on bulk band structures, which neglect the interface dipole moment, therefore, possess significant deviations.

**Conclusion**

In summary, we calculated the band gap and SO splitting of GaAs$_{1-x-y}$P$_y$Bi$_x$ and the band offsets of GaAs/GaAs$_{1-x-y}$P$_y$Bi$_x$ heterojunctions using DFT. Fitted analytic expressions of the band gap and SO splitting dependences on compositions were obtained, which are consistent with available experimental trends. The band gap changes were found to be primarily contributed by the local distortions around P and Bi atoms. A composition requirement for GaAs$_{1-x-y}$P$_y$Bi$_x$ to achieve lower Auger recombination ratio than GaAs was obtained. Finally, we found that GaAs/GaAs$_{1-x-y}$P$_y$Bi$_x$ heterojunctions with achievable compositions and small lattice mismatch can exhibit a type-I band diagram and the band offsets can be tuned to suit the requirements of infrared LED and LD applications.

**Supplementary Material**

See supplementary material for the (a) procedure to obtain band offsets, (b) linear correction to DFT band gaps, (c) fitting errors of Eqn. 1, 2, 4, and 5, (d) reasoning of the strain effects on band gap, (e) projected density of states of GaAs$_{0.5}$P$_{0.398}$Bi$_{0.102}$, (f) comparison between Eqn. 1b, 1c, and 4b with experimental results, and (g) complete list of the band gap, SO splitting, and lattice length.

**Acknowledgment**

This research was primarily supported by NSF through the University of Wisconsin Materials Research Science and Engineering Center (Grant No. DMR-1121288). The authors gratefully acknowledge the use of clusters supported by NSF through the University of Wisconsin Materials Research Science and Engineering Center (Grant No. DMR-1121288). Computing resources in this work also benefited from the use of the Extreme Science and Engineering Discovery Environment (XSEDE), which is supported by National Science Foundation Grant No. ACI-1053575 and the compute resources and assistance of the UW-Madison Center For High Throughput Computing (CHTC) in the Department of Computer Sciences.

**Reference**

[1] D. S. Meyaard, G. B. Lin, J. Cho, E. F. Schubert, H. Shim, S. H. Han, M. H. Kim, C. Sone, and Y. S. Kim, Appl. Phys. Lett. **102** (25), 251114 (2013).
[2] J. Piprek, Appl. Phys. Lett. **107** (3), 031101 (2015).
[3] Y. Bai, N. Bandyopadhyay, S. Tsao, S. Slivken, and M. Razeghi, Appl. Phys. Lett. **98** (18), 181102 (2011).
[4] L. Shterengas, R. Liang, G. Kipshidze, T. Hosoda, G. Belenky, S. S. Bowman, and R. L. Tober, Appl. Phys. Lett. **105** (16), 161112 (2014).
[5] J. Piprek, Phys. Status Solidi A **207** (10), 2217 (2010).
[6] J. Cho, E. F. Schubert, and J. K. Kim, Laser Photonics Rev. **7** (3), 408 (2013).
[7] L. A. Coldren, S. W. Corzine, and Milan Mashanovitch, *Diode lasers and photonic integrated circuits*, 2nd ed. (Wiley, Hoboken, N.J., 2012).
[8] J. Iveland, L. Martinelli, J. Peretti, J. S. Speck, and C. Weisbuch, Phys. Rev. Lett. **110** (17), 177406 (2013).
[9] K. A. Bulashevich and S. Y. Karpov, Phys. Status Solidi C **5** (6), 2066 (2008).
[10] Handong Li, *Bismuth-containing compounds*. (Springer, New York, 2013).
[11] Z. Batool, K. Hild, T. J. C. Hosea, X. Lu, T. Tiedje, and S. J. Sweeney, J. Appl. Phys. **111** (11), 113108 (2012).
[12] D. Steiauf, E. Kioupakis, and C. G. Van de Walle, ACS Photonics **1** (8), 643 (2014).
[13] G. Vardar, S. W. Paleg, M. V. Warren, M. Kang, S. Jeon, and R. S. Goldman, Appl. Phys. Lett. **102** (4), 042106 (2013).
[14] J. Puustinen, M. Wu, E. Luna, A. Schramm, P. Laukkanen, M. Laitinen, T. Sajavaara, and M. Guina, J. Appl. Phys. **114** (24), 243504 (2013).
[15] V. Pacebutas, R. Butkute, B. Cechavicius, J. Kavaliauskas, and A. Krotkus, Thin Solid Films **520** (20), 6415 (2012).
[16] H. Jacobsen, B. Puchala, T. F. Kuech, and D. Morgan, Phys. Rev. B **86** (8), 085207 (2012).
[17] G. F. Luo, S. J. Yang, J. C. Li, M. Arjmand, I. Szlufarska, A. S. Brown, T. F. Kuech, and D. Morgan, Phys. Rev. B **92** (3), 035415 (2015).
[18] K. Forghani, Y. X. Guan, M. Losurdo, G. F. Luo, D. Morgan, S. E. Babcock, A. S. Brown, L. J. Mawst, and T. F. Kuech, Appl. Phys. Lett. **105** (11), 111101 (2014).
[19] G. Kresse and J. Furthmuller, Phys. Rev. B **54** (16), 11169 (1996).
[20] G. Kresse and J. Furthmuller, Comp. Mater. Sci. **6** (1), 15 (1996).
[21] A. van de Walle, P. Tiwary, M. de Jong, D. L. Olmsted, M. Asta, A. Dick, D. Shin, Y. Wang, L. Q. Chen, and Z. K. Liu, Calphad **42**, 13 (2013).

[22] H. M. Al-Allak and S. J. Clark, Phys. Rev. B **63** (3), 033311 (2001).
[23] A. Janotti, S. H. Wei, and S. B. Zhang, Phys. Rev. B **65** (11), 115203 (2002).
[24] H. W. King, J. Mater. Sci. **1**, 79 (1966).
[25] B. Cordero, V. Gomez, A. E. Platero-Prats, M. Reves, J. Echeverria, E. Cremades, F. Barragan, and S. Alvarez, Dalton Trans. (21), 2832 (2008).
[26] K. Alberi, J. Wu, W. Walukiewicz, K. M. Yu, O. D. Dubon, S. P. Watkins, C. X. Wang, X. Liu, Y. J. Cho, and J. Furdyna, Phys. Rev. B **75** (4), 045203 (2007).
[27] I. Vurgaftman, J. R. Meyer, and L. R. Ram-Mohan, J. Appl. Phys. **89** (11), 5815 (2001).
[28] W. Huang, K. Oe, G. Feng, and M. Yoshimoto, J. Appl. Phys. **98** (5), 053505 (2005).
[29] S. Francoeur, M. J. Seong, A. Mascarenhas, S. Tixier, M. Adamcyk, and T. Tiedje, Appl. Phys. Lett. **82** (22), 3874 (2003).
[30] K. Alberi, O. D. Dubon, W. Walukiewicz, K. M. Yu, K. Bertulis, and A. Krotkus, Appl. Phys. Lett. **91** (5), 051909 (2007).
[31] B. Fluegel, S. Francoeur, A. Mascarenhas, S. Tixier, E. C. Young, and T. Tiedje, Phys. Rev. Lett. **97** (6), 067205 (2006).
[32] D. V. Morgan and Robin H. Williams, *Physics and Technology of Heterojunction Devices*. (Peter Peregrinus Ltd., London, United Kindoms, 1991), p.241.
[33] M. Usman, C. A. Broderick, Z. Batool, K. Hild, T. J. C. Hosea, S. J. Sweeney, and E. P. O'Reilly, Phys. Rev. B **87** (11), 115104 (2013).
[34] A. Baldereschi, S. Baroni, and R. Resta, Phys. Rev. Lett. **61** (6), 734 (1988).

# Supplementary Material for "First-principles Predications of Electronic Properties of GaAs$_{1-x-y}$P$_y$Bi$_x$ and GaAs$_{1-x-y}$P$_y$Bi$_x$–based Heterojunctions"

Guangfu Luo, Kamran Forghani, Thomas F. Kuech, and Dane Morgan*
*E-mail: ddmorgan@wisc.edu

## I. A three-step procedure to obtain band offsets

Band offsets of the GaAs/GaAs$_{1-x-y}$P$_y$Bi$_x$ heterojunction are calculated based on positions of valance band maximum (VBM) and conduction band minimum (CBM) relative to macroscopic electrostatic potential (MEP) with a three-step procedure.[1] First, the MEP along [001] of the heterojunction is calculated. To reduce the fluctuations of using a finite-sized supercell with a specific P-Bi order, the average MEP of several heterojunctions with different P and Bi SQS configurations is used. Second, the relative position of VBM in the heterojunction is determined according to the VBM positions relative to MEP in the GaAs and GaAs$_{1-x-y}$P$_y$Bi$_x$ bulk. Then, the VBM band offset, $\Delta E_{VBM}$, can be obtained. Third, the CBM band offset, $\Delta E_{CBM}$, is calculated according to the $\Delta E_{VBM}$ and band gap difference, $\Delta E_g$, between GaAs and GaAs$_{1-x-y}$P$_y$Bi$_x$ bulk, namely, $\Delta E_{CBM} = \Delta E_g - \Delta E_{VBM}$. This approach implies that the correction of LDA band gap is modeled as a shift of conduction bands. The aforementioned method has been applied to GaAs/GaAsP and GaAs/GaAsBi heterojunctions and gave band offsets consistent with previous experiments.[2]

## II. Correction to theoretical band gaps

Figure S1 compares the theoretical band gaps of GaAs, GaAs$_{1-x}$Bi$_x$, and GaAs$_{1-y}$P$_y$ with available experimental results.[3-7] As expected, the band gaps calculated with LDA functional and spin-orbit coupling (LDA+SOC) are significantly underestimated (orange squares). However, a linear correction of the DFT values by 0.88 eV + 1.13 $E^g$ (red diamonds) greatly remedies the underestimation and standard deviation relative to experimental values is reduced to 0.04 eV.

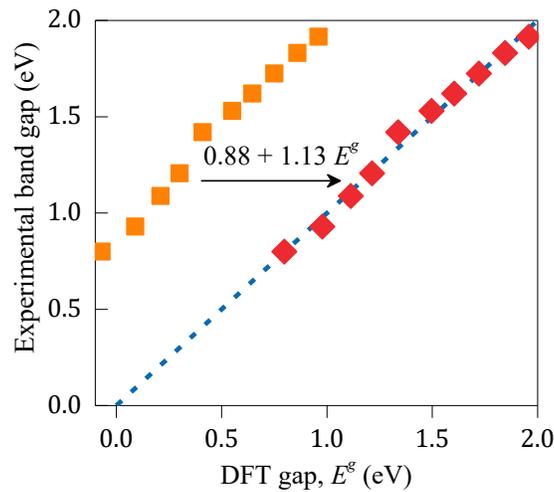

FIG. S1. DFT band gaps with and without a linear correction versus experimental values for fully relaxed GaAs, GaAs$_{1-x}$Bi$_x$ ($x$ = 0.028, 0.046, 0.074, and 0.102), and GaAs$_{1-y}$P$_y$ ($y$ = 0.093, 0.167, 0.250,

0.333, and 0.398). The experimental band gaps of relaxed GaAs$_{1-x}$Bi$_x$ and GaAs$_{1-y}$P$_y$ are obtained from previous widely-accepted fittings by Ref.3,7 and Ref.4, respectively.

### III. Band gap and SO splitting of GaAs$_{1-x-y}$P$_y$Bi$_x$

The complete list of band gap and SO splitting of fully relaxed and strained GaAs$_{1-x-y}$P$_y$Bi$_x$ is provided in a separate Excel file. Figure S2 compares the DFT-based values and the fitted values using Eqn. 1, 2, 4, and 5 in the main text. The fittings are generally good with standard deviation of 0.03 eV or 0.01 eV between the fitted and DFT calculated band gap or SO splitting.

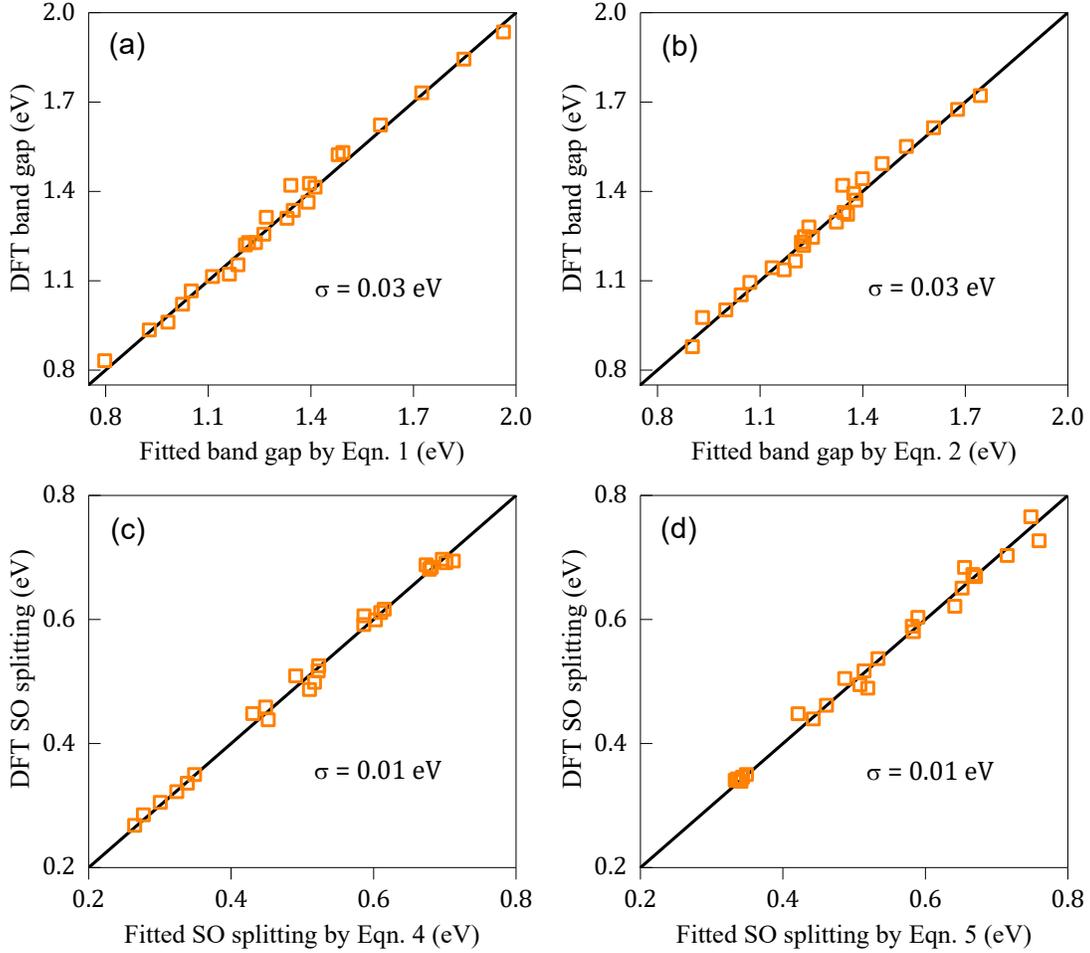

FIG. S2. Fitted values versus DFT-based values for (a) band gap of fully relaxed GaAs$_{1-x-y}$P$_y$Bi$_x$, (b) band gap of strained GaAs$_{1-x-y}$P$_y$Bi$_x$, (c) SO splitting of fully relaxed GaAs$_{1-x-y}$P$_y$Bi$_x$, and (d) SO splitting of strained GaAs$_{1-x-y}$P$_y$Bi$_x$. The standard deviation of each fitting is indicated.

### IV. Analytic expression of strain effects on band gap

Here we derive a relationship between the band gap change of GaAs$_{1-x-y}$P$_y$Bi$_x$ strained to a (001) plane relative to its fully relaxed counterpart. Generally, the band gap $E^g$ of a IV-IV, III-V, and II-VI zinc-blend semiconductor or insulator is linearly proportional to its bond energy,[8] which is inversely proportional to bond length $l$,[9] namely, $E^g \propto 1/l$. Therefore, the relative change of $E^g$ follows Eqn. S1.

$$\frac{\Delta E^g}{E^g} \propto -\frac{\Delta l}{l} \tag{S1}$$

The bond length change, $\Delta l$, can be calculated from the lattice changes in plane (001), $\Delta a$, and the lattice change along [001] direction, $\Delta c$, both of which are connected by the strain-stress relationship of a compound. If we approximate the mechanical properties of $GaAs_{1-x-y}P_yBi_x$ as those of GaAs,[10] as shown in Eqn. S2, and consider a symmetrical strain in (001) plane, namely $\sigma_{xx} = \sigma_{yy} = 1$ and all other $\sigma_{ij}= 0$, the relative strain changes along [100] and [001] has the relationship of $\varepsilon_{zz} \approx -0.9\ \varepsilon_{xx}$.

$$\begin{pmatrix} \sigma_{xx} \\ \sigma_{yy} \\ \sigma_{zz} \\ \sigma_{yz} \\ \sigma_{xz} \\ \sigma_{xy} \end{pmatrix} = \begin{pmatrix} C_{11} & C_{12} & C_{12} & 0 & 0 & 0 \\ C_{12} & C_{11} & C_{12} & 0 & 0 & 0 \\ C_{12} & C_{12} & C_{11} & 0 & 0 & 0 \\ 0 & 0 & 0 & C_{44} & 0 & 0 \\ 0 & 0 & 0 & 0 & C_{44} & 0 \\ 0 & 0 & 0 & 0 & 0 & C_{44} \end{pmatrix} \cdot \begin{pmatrix} \varepsilon_{xx} \\ \varepsilon_{yy} \\ \varepsilon_{zz} \\ 2\varepsilon_{yz} \\ 2\varepsilon_{xz} \\ 2\varepsilon_{xy} \end{pmatrix} \quad (S2)$$

$C_{11} = 118.80\ \text{GPa}, C_{12} = 53.70\ \text{GPa}, C_{44} = 59.40\ \text{GPa}$

Therefore, $\Delta c$ and $\Delta a$ follows Eqn. S3.

$$\Delta c \approx -0.9\ \Delta a \quad (S3)$$

Based on Eqn. S3 and the geometry of zinc-blende structure, $\Delta l/l$ is approximated as Eqn. S4.

$$\frac{\Delta l}{l} \approx \left( \sin\theta^2 + \frac{-0.9}{\sqrt{2}} \sin\theta \cos\theta \right) \frac{\Delta a}{a}, \theta = 54.74°$$

$$\frac{\Delta l}{l} \approx 0.37 \frac{\Delta a}{a} \quad (S4)$$

Combining Eqn. S1 and S4 leads to Eqn. S5.

$$\frac{\Delta E^g}{E^g} \propto \frac{\Delta a}{a} \quad (S5)$$

To determine the unknown coefficient in Eqn.S5, we fit the DFT data of $GaAs_{1-x-y}P_yBi_x$ strained on GaAs(001) as shown in Fig. S3 and obtained Eqn. S6:

$$\frac{\Delta E^g}{E^g} \approx -6.96 \frac{\Delta a}{a} \quad (S6)$$

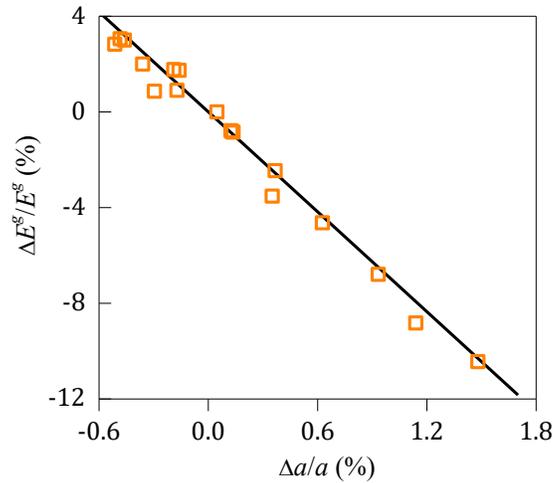

FIG. S3. Relationship between $\Delta E^g/E^g$ and $\Delta a/a$ for DFT-based results of fully relaxed and strained $GaAs_{1-x-y}P_yBi_x$. The slope of the fitting line is -6.96 and the standard error of this slope is 0.24.

## V.  Projected density of states of $GaAs_{0.5}P_{0.398}Bi_{0.102}$

Figure S4 shows that the states from Bi and P spread throughout the whole bands, rather than localize near the band edges as predicted by the valence band anti-crossing model.

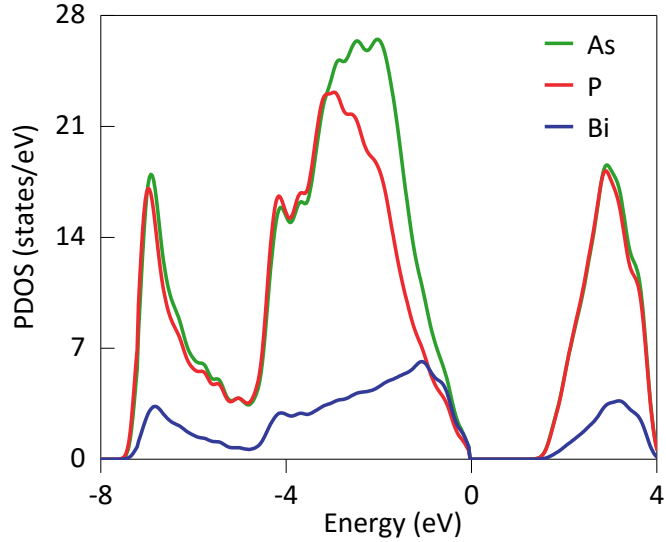

FIG. S4. Projected density of sates (PDOS) of As, P, and Bi in fully relaxed $GaAs_{0.5}P_{0.398}Bi_{0.102}$.

## VI.  Comparison between theoretical and experimental results of $GaAs_{1-x}Bi_x$ and $GaAs_{1-y}P_y$

Figure S5 compares our theoretical fitting results with the experimental band gap[4] of $GaAs_{1-y}P_y$ and band gap[3,5-7] and SO splitting[3,7,11] of $GaAs_{1-x}Bi_x$. In the examined ranges, the largest differences between the experiment-based data and Eqn. 1b, Eqn. 1c, Eqn. 4b are about 0.03, 0.02, and 0.08 eV, respectively.

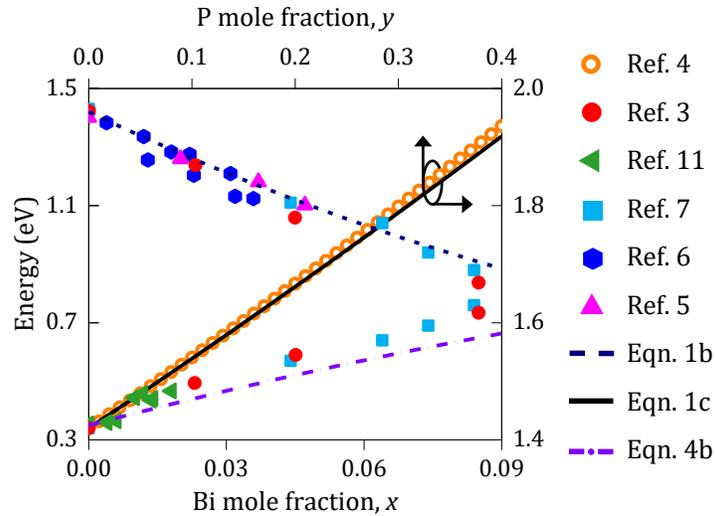

FIG. S5. Comparison between experimental data or best fitting of experimental data and our equations for band gap (upper branch) and SO splitting (lower branch) of $GaAs_{1-x}Bi_x$ and band gap of $GaAs_{1-y}P_y$ (top and right axes).

# Reference

1. H. M. Al-Allak and S. J. Clark, Phys. Rev. B **63** (3), 033311 (2001).
2. H. Kim, K. Forghani, Y. Guan, G. Luo, A. Anand, D. Morgan, T. F. Kuech, L. J. Mawst, Z. R. Lingley, B. J. Foran, and Y. Sin, Semicond. Sci. Technol. **30** (9), 094011 (2015).
3. Z. Batool, K. Hild, T. J. C. Hosea, X. Lu, T. Tiedje, and S. J. Sweeney, J. Appl. Phys. **111** (11), 113108 (2012).
4. I. Vurgaftman, J. R. Meyer, and L. R. Ram-Mohan, J. Appl. Phys. **89** (11), 5815 (2001).
5. W. Huang, K. Oe, G. Feng, and M. Yoshimoto, J. Appl. Phys. **98** (5), 053505 (2005).
6. S. Francoeur, M. J. Seong, A. Mascarenhas, S. Tixier, M. Adamcyk, and T. Tiedje, Appl. Phys. Lett. **82** (22), 3874 (2003).
7. K. Alberi, O. D. Dubon, W. Walukiewicz, K. M. Yu, K. Bertulis, and A. Krotkus, Appl. Phys. Lett. **91** (5), 051909 (2007).
8. P. Manca, Journal of the Physics and Chemistry of Solids **20** (3-4), 268 (1961).
9. L. Pauling, J. Phys. Chem. **58**, 662 (1954).
10. Allan F. Bower, *Applied mechanics of solids*. (CRC Press, Taylor & Francis Group, 2010), p.88.
11. B. Fluegel, S. Francoeur, A. Mascarenhas, S. Tixier, E. C. Young, and T. Tiedje, Phys. Rev. Lett. **97** (6), 067205 (2006).